\documentstyle[11pt,aaspp4,flushrt,epsfig]{article}
\renewcommand{\vec}[1]{\mbox{\boldmath $\displaystyle #1$}}
\newcommand{\grad}{\vec{\nabla}}
\newcommand{\divr}{{\rm div}\,}
\newcommand{\curl}{{\rm curl}\,}

\newcommand{\vcross}{\mbox{\boldmath $\times$}}
\newcommand{\ee}[2]{#1 \times 10^{#2}}
\newcommand{\K}{\,{\rm K}}
\newcommand{\cps}{\,{\rm c\,s^{-1}}}
\newcommand{\gram}{\,{\rm g}}
\newcommand{\secd}{\,{\rm s}}
\newcommand{\yr}{\,{\rm yr}}
\newcommand{\cm}{\,{\rm cm}}
\newcommand{\km}{\,{\rm km}}
\newcommand{\gauss}{\,{\rm G}}
\newcommand{\erg}{\,{\rm erg}}
\newcommand{\MeV}{\,{\rm MeV}}
\newcommand{\keV}{\,{\rm keV}}
\newcommand{\cpi}{\varpi}
\newcommand{\msun}{M_{\odot}}
\begin{document}
\title{
Thermonuclear Burning on the Accreting X-Ray Pulsar  GRO~J1744-28
}
\author{
Lars Bildsten and Edward F.\ Brown
}
\affil{
Department of Physics and Department of Astronomy\\
University of California, Berkeley, CA 94720\\
I: bildsten@fire.berkeley.edu, ebrown@astron.berkeley.edu
}
\authoremail{
bildsten@fire.berkeley.edu, ebrown@astron.berkeley.edu
}
\begin{abstract}

We investigate the thermal stability of nuclear burning on the
accreting X-ray pulsar GRO J1744-28. The neutron star's dipolar
magnetic field is $\lesssim 3\times 10^{11} \gauss$ if persistent
spin-up implies that the magnetospheric radius is less than the
co-rotation radius. Bildsten earlier noted that magnetic fields this
weak might not quench the vigorous convection sometimes associated
with thermonuclear instabilities, so that a convective burning front
can propagate around the star in a few seconds and rapidly release the
accumulated nuclear energy.  After inferring the properties of the
neutron star, we study the thermal stability of hydrogen/helium
burning and show that thermonuclear instabilities are unlikely causes
of the hourly bursts seen at very high accretion rates. We then
discuss how the stability of the thermonuclear burning depends on both
the global accretion rate and the neutron star's magnetic field
strength.  We emphasize that the appearance of the instability (i.e.,
whether it looks like a Type I X-ray burst or a flare lasting a few
minutes) will yield crucial information on the neutron star's surface
magnetic field and the role of magnetic fields in convection.

We suggest that a thermal instability in the accretion disk is the
origin of the long ($\sim 300$ days) outburst and that the recurrence
time of these outbursts is $>50$ years. We also discuss the nature of
the binary and point out that a velocity measurement of the stellar
companion (most likely a Roche-lobe filling giant with $m_K\gtrsim
17$) will constrain the neutron star mass.

\end{abstract}

\begin{center}
\bf To Appear in the Astrophysical Journal
\end{center}

\section{
Introduction
}

The transient accretion-powered pulsar ($\nu=2.14$ Hz) GRO~J1744-28
was discovered by the Burst and Transient Source Experiment ({\em
BATSE\/}) aboard the Compton Gamma Ray Observatory ({\em CGRO\/})
during a day of rapid bursting (about twenty bursts per hour) on 1995
December 2 (Kouveliotou et al.\ 1996a; Finger et al.~1996a).  Finger
et al.\ (1996a) measured the 11.8 day orbit and found that GRO
J1744-28 was spinning up at a rate $\dot \nu=(3.5$--$12.2)\times
10^{-12} \secd^{-2}$ between 1995 December 15 and 1996 January 23.
The pulsar subsequently settled into a regime of hourly bursting with
burst durations of 2--7 seconds, as seen by the PCA instrument on the
Rossi X-Ray Timing Explorer ({\em RXTE\/}) (Swank 1996; Giles et al.\
1996) and the OSSE instrument aboard {\em CGRO\/} (Strickman et al.\
1996). A qualitatively similar burst has been seen from another
accreting pulsar (see our discussion of SMC X-1 in \S2), but
recurrent bursts of this nature have never been observed.

The bursting behavior has two obvious energy sources: accretion and
thermonuclear burning. Lewin et al.\ (1996) presented the case for
accretion-powered bursts via analogy to the Rapid Burster.  The rapid
bursts (one every few minutes) are responsible for up to 50\% of the
total time averaged luminosity above 20 keV and thus cannot have a
thermonuclear origin (Kouveliotou et al.\ 1996a). The hourly bursts
(over 3000 observed by {\em BATSE\/} [Kouveliotou et al.\ 1996c]) are
responsible for a much smaller fraction of the total luminosity, at
least above 20 keV. The average burst fluence was $7\times 10^{-7}
\erg\cm^{-2}$ (20--50 keV) on 1996 January 15, when there were 40 per
day and the persistent luminosity was 2.5 Crab in the 20--100 keV band
(1996 January 16 [Fishman et al.\ 1996]) and $4.4\pm 0.3$ Crab in the
8--20 keV band (1996 January 14--1996 January 15 [Sazonov \& Sunyaev
1996]). Above 20 keV, this gives a time-averaged burst luminosity
60--100 times smaller than the steady-state accretion luminosity.  The
burst energetics below 20 keV (where most of the energy is emitted) is
still uncertain due to dead-time corrections in the PCA. Preliminary
indications are that these corrections are large enough so that the
bursts might be responsible for more than 5\% of the $<20$ keV
emission (Jahoda 1996, private communication).

The burst energies were marginally consistent with nuclear energy
release and motivated our theoretical study of thermonuclear burning
on this unusual X-ray pulsar. Additional motivations were the
matchings of the characteristic decay time (2--10 seconds [Strickman
et al.\ 1996]) with the cooling time at the nuclear burning depth and
the mean recurrence time (about 30 minutes for the 260 bursts seen by
OSSE from 1996 January 16 to 1996 January 30 [Strickman et al.\ 1996])
with the time to accumulate enough fuel for an instability. The
primary observational evidence against {\em all\/} of these bursts
having a thermonuclear origin is (1) the independence of the
recurrence time from the accretion rate (at least when it is brighter
than 200--400 mCrab [Giles et al.\ 1996]), (2) the existence of
``foreshocks'' before the bursts (Giles et al.\ 1996), (3) the lack of
spectral evolution during the burst, and (4) the global burst
energetics from the PCA instrument (if the dead-time corrections are
understood).

We consider here the possibility that this X-ray pulsar has an
unusually low field and therefore might exhibit different
thermonuclear burning behavior than conventional X-ray pulsars.  It is
unlikely that the thermonuclear burning was unstable at the peak of
the outburst. Hence we concentrate on lower accretion rates, for which
the burning will most likely be thermally unstable and will manifest
itself as Type I X-ray bursts or flares of a few minutes duration.

Since the presence and character of a thermonuclear instability
depends on the neutron star's magnetic field ($B$) and global
accretion rate ($\dot M$), we begin in \S2 by summarizing the
indirect inferences about these quantities. The spin behavior points to
an especially weak ($\ll 10^{12} \gauss$) dipolar magnetic field
component and constrains the global accretion rate as well. We also
compare GRO~J1744-28's bursting behavior at the peak of the outburst
to that of SMC X-1, for which similar arguments point to a low dipole
field. In \S3, we constrain the binary properties, infer a
time-averaged mass transfer rate, and speculate on the origin of the
long-term transient behavior. The thermonuclear stability of the
accreted hydrogen and helium is discussed in \S4 for a range of $\dot
M$'s.  Section 5 is a discussion of the magnetic field's role in the
stability and character of nuclear burning in accreting X-ray
pulsars. In particular, we discuss in detail how the star will behave
as $\dot M$ decreases.  We conclude, in \S6, by describing what a
successful identification of a thermonuclear instability implies about
the neutron star's properties.

\section{ 
Properties of the Accreting Neutron Star and Comparison to SMC X-1
}

In addition to the bursting behavior, this pulsar is unusual because
of its unusually high spin frequency and its steady spin-up over a
large range of accretion rates. In the context of magnetic accretion,
these facts imply a dipole field lower than most other accreting
pulsars.  The magnetosphere is located at $r_m=\xi r_A$ (Ghosh \& Lamb
1979), where $r_A$ is a characteristic length found by equating
magnetic and fluid stresses, and $\xi$ is a model-dependent
dimensionless number.  Estimates of $\xi$ range from $\approx 0.52$
(Ghosh \& Lamb 1979) to $\approx 1$ (Arons 1993; Ostriker \& Shu 1995;
Wang 1995). A measurement of $\xi$ obtained from the observed quasi-periodic
oscillations in the accreting pulsar A0535+26 (Finger, Wilson, \&
Harmon 1996b) gave $\xi\approx 1$. Since the neutron star is spinning
up, the magnetospheric radius, $r_m$, must be less than the
co-rotation radius, $r_{\rm co}$, which implies an {\em upper\/} limit
on the surface strength of the dipolar componenet of the magnetic
field
\begin{equation}\label{eq:rm<rco}
B\lesssim 3\times 10^{11}\gauss\;\xi^{-7/4}
\left(\frac{\dot{M}}{\ee{8}{-9} \msun \yr^{-1}}\right)^{1/2}
\left(\frac{10\km}{R}\right)^3,
\end{equation}
in agreement with previous estimates (Finger et al.~1996a; Sturner \&
Dermer 1996). The magnetospheric radius is presently unknown;
continual spin-up at lower $\dot M$'s will reduce this upper
limit. This upper bound is less than the typical X-ray pulsar field
strength, which, as we discuss in \S5, changes the nature of the
convection; GRO~J1744-28 is the first high accretion rate X-ray pulsar
beneath this limit.

The X-ray spectrum above 20 keV falls very steeply (Kouveliotou et
al.\ 1996a; Strickman et al.\ 1996) and is consistent with being above
the characteristic cut-off energy found by {\em RXTE}/PCA ($\approx
15\mbox{--}20\keV$ [Swank 1996; Giles et al.\ 1996]). Daumerie et al.\
(1996) argue that this spectrum and the increase in pulse fraction
with energy imply a surface field $\sim 10^{12} \gauss$. This argument
is at odds with our upper limit on the dipolar component (for $\xi=1$)
and might imply that higher order magnetic moments are present.

We use the measured spin-up rate $\dot \nu$ to constrain the accretion
rate onto the neutron star. The maximum specific angular momentum of
the accreted matter is $l_{\rm max}\equiv(GM_xr_{\rm co})^{1/2}$,
where $r_{\rm co}=1.0\times 10^8 \cm\approx 100 R$ is the co-rotation
radius (where the Kepler frequency equals the neutron star's spin
frequency) for a $M_x=1.4\msun$ neutron star. Because the observed
torque $2\pi I \dot \nu$ must be less than $\dot M l_{\rm max}$ (Ghosh
\& Lamb 1979; Chakrabarty et al.\ 1993), we set a lower bound to
$\dot M$,
\begin{equation}
\dot M > 8\times 10^{-9} \msun\yr^{-1}
\left(\frac{\dot{\nu}}{10^{-11}\secd^{-2}}\right)
\left(\frac{R}{10\km}\right)^2,
\end{equation}
where $I=0.4 M_x R^2$ is the neutron star's moment of inertia (a good
approximation for our choice of mass and radius [Ravenhall \& Pethick
1994]). The spherically averaged local accretion rate,
\begin{equation}
\dot{m}_{\rm sph} \equiv \frac{\dot{M}}{4\pi R^2} > \ee{4}{4}
\gram\cm^{-2}\secd^{-1}
\left(\frac{\dot{\nu}}{10^{-11} \secd^{-2}}\right),
\end{equation}
is then {\em independent\/} of the neutron star's radius. Using a rough
bolometric flux for 1996 January 15 of $10^{-7} \erg \cm^{-2}
\secd^{-1}$, we infer a minimum distance $d=3\,{\rm kpc}$. As we 
note later, the extinction and position in the galaxy most likely
place the object at least 8 kpc away.

The only other accreting pulsar which has always been spinning up
($\dot \nu=2.4\times 10^{-11} \secd^{-2}$) and for which our earlier
arguments yield a comparable field strength is SMC X-1 ($\nu=1.4 \,
{\rm Hz}$). For a typical accretion rate $\dot M\approx 4\times
10^{-8} \msun \yr^{-1}$ (Levine et al.\ 1993), we infer $B<10^{12}
\gauss$. Angelini, Stella, and White (1991) discovered an X-ray burst
from SMC X-1 during an {\em EXOSAT\/} observation on 18 October 1984
when the 1-16 keV luminosity was $L\approx 4\times 10^{38} \erg
\secd^{-1}$. This burst is {\em very\/} similar to the large hourly
bursts seen from GRO~J1744-28. The burst rose by a factor of three
within one second, lasted for about 80 seconds, and was followed by a
35\% decline in the persistent flux. There was no evidence for
spectral changes during or after the burst, and the pulse fraction and
phase remained constant. The recurrence time must be long, as only one
burst was seen in $\approx 20$ hours of observation. Angelini et al.\
(1991) noted the similarities to the Rapid Burster and argued for
accretion as the energy source for these events. They also noted that
the variability of the source has a strong underlying quasi-period of
a few minutes, which was present in all {\em EXOSAT\/} observations
except for the five hours following the burst. About 10\% of the total
luminosity of the source is in these variations.

\section{
Properties of the Binary
}

Following the {\em ROSAT\/} positioning of this object (Kouveliotou et
al.\ 1996b), Augusteijn et al.\ (1996) identified the variable
infra-red counterpart, which was present in an earlier image (8
February 1996) of Blanco, Lidman, and Glazebrook (1996) at
$m_K=15.7\pm 0.3$ and was undetected and at least a magnitude fainter
on 28 March 1996. This light is most likely X-rays reprocessed by
either the accretion disk or companion (Augusteijn et al.\ 1996).  The
most obvious Roche-lobe filling object is a first-ascent red giant
branch star with a degenerate helium core of mass $M_{\rm He}$ and an
overlying hydrogen envelope (Finger et al.~1996a; Sturner \& Dermer
1996).  Hydrogen shell burning via the CNO cycle supplies a luminosity
strongly dependent on only the helium core mass, which allows us to
estimate the stellar luminosity and expected IR magnitude.  In the
absence of mass or angular momentum loss, this binary evolves due to
the expansion of the red giant as the helium core mass grows.

\subsection{
Constraints on the Optical Companion and Neutron Star Mass 
}

We solve for the companion mass, $M_c$, by using the orbital
parameters $P_{\rm orb}=11.83 \,{\rm d}$ and $a_x \sin i=2.63
\,{\rm lt\mbox{-}sec}$ (Finger et al.~1996a) and the
core-mass/luminosity relations of Webbink, Rappaport, \& Savonije
(1983).  We presume that the giant fills the Roche lobe estimated by
Eggleton (1983) and fix $M_x=1.4 \msun$. For metallicities
$Z=0.02(0.0001)$, the first allowed solution (corresponding to an
unrealistic hydrogen envelope mass of zero) has $M_c/\msun =
0.216\,(0.232)$ and inclinations less than 18--19 degrees ($\cos
i>0.95$).  The solution in the middle of the allowed range (i.e.,
$\cos i=0.975$) has $M_c=0.334\,\msun$, $M_{\rm He}=0.22\,(0.24)
\msun$, $L\approx 12 L_\odot$, an orbital separation of $26
R_\odot$, a stellar radius of $6.9R_\odot$, and a projected companion
velocity $K\approx 20 \km\secd^{-1}$.  The helium core mass is
slightly larger for the metal-poor case to compensate for fewer
catalysts. The strong dependence of the giant radius on $M_{\rm He}$
makes the inferred core mass (shown by the dotted line in Figure 1)
nearly independent of the inclination angle and always close to
$0.22\, (0.24) \msun$. The dashed line in Figure 1 shows $M_c$ as a
function of $\cos i$ for $Z=0.02$ and $M_x=1.4\msun$.  The resulting
mass transfer rate (Webbink et al.\ 1983) is given by the solid line
in Figure 1.  For the ``typical'' solution presented above, the
average mass transfer rate is $\dot{M}\approx 10^{-9} \msun \yr^{-1}$,
and the hydrogen envelope mass is $0.1 \msun$. This implies a
lifetime, nearly independent of the metallicity, of $10^8 \yr$. Most
of the giant's envelope goes onto the neutron star (for conservative
evolution).

Unless we are looking at the system nearly pole-on, we must conclude
that $M_c\approx 0.3 \msun$. If this low-mass star followed a normal
evolutionary track, then it must have started mass transfer as a
$\approx \msun$ star. If all the departed mass accreted onto the
compact object, then either the neutron star is more massive than its
``birthweight'' ($1.2$--$1.7 \msun$ [Timmes, Woosley, \& Weaver
1996]), or the neutron star was a white dwarf that underwent
accretion-induced collapse. Considering more massive neutron stars
does not alleviate the companion's substantial mass loss but does
increase the inclination angle (e.g., $\cos i=0.9$ for $M_x=2.0
\msun$, a doubling of the allowed phase space).

Depending on the inclination angle, an infrared measurement of the
orbit of this system might actually constrain the neutron star
mass. The projected companion velocity is 
\begin{equation}
K = 	\left(\frac{2\pi a_x \sin i}{P_{\rm orb}}\right)
	\left(\frac{M_x}{M_c}\right) = 4.85 \km\secd^{-1}
	\left(\frac{M_x}{M_c}\right).
\end{equation}
We find the minimum companion mass by requiring (1) that the star be
luminous enough so that it fills the Roche lobe, and (2) that the
overlying hydrogen envelope have a minimum mass of $10^{-2}
\msun$ so as to live for at least $10^{7} \yr$ and to be fully
convective. The minimum companion mass ($M_c=0.22 \msun$) is then
basically independent of $M_x$, so that the maximum $K$ for the
Roche-lobe filling giant hypothesis is $K_{\rm max}=21 \km
\secd^{-1}(M_x/\msun)$. Hence, a measurement of $K$ can 
potentially constrain the neutron star mass. For example, if $M_x=2
\msun$ and we presume that $\cos i$ is uniformly distributed between
0.9 and 1.0, then there is a 50\% chance of measuring a velocity
larger than the maximum for the $1.4 \msun$ neutron star case.  This
will be a tough job. We find that the stellar effective temperature is
in the range 4000--4200$\K$, which, for the luminosity derived above,
gives $M_K\approx -0.2$ (Bessell \& Brett 1988). If the extinction is
$A_V\approx 25$, as implied by the $N_H\approx (5$--$6)\times 10^{22}
\cm^{-2}$ measurement of Dotani et al.\ (1996), then $A_K\approx 3$ and for
a distance of 8 kpc, we expect $m_K\approx 17$.

\subsection{
The Long-Term Accretion History
}

Consistent with the transient nature of this binary, the inferred
long-term mass transfer rate is less than the current accretion rate.
This possibly indicates that the present outburst (which so far has
lasted for over 300 days) arises from a thermal instability in the
accretion disk similar to that occurring in dwarf novae. Indeed, the
implied long term mass transfer rate is much less than the amount
needed for stable mass transfer (Shafter 1992). Our orbital
calculations give the distance from the neutron star to where the
accreting matter strikes the accretion disk as $R_r\approx 4 R_\odot$
(Lubow \& Shu 1975). The thermal instability occurs when the column
density of matter ($\Sigma$) accumulated at this radius exceeds
$\gtrsim 10^3 \gram \cm^{-2}$ (Cannizzo \& Wheeler 1984), in which
case the accumulated mass is $\gtrsim R_r^2 \Sigma \sim \ee{5}{-8}
\msun $. This requires over 50 years of mass transfer from the giant
and supplies more than enough material to power the large outburst.

The accretion will most likely be halted by a propeller effect at low
accretion rates (Illarionov \& Sunyaev 1975).  However, even when not
accreting, the neutron star has a luminosity from the cooling core (at
temperature $T_c$) given by Gudmundsson, Pethick, and Epstein (1983) as
\begin{equation}\label{cool} 
L_{\rm core} = \ee{6}{32}\erg\secd^{-1}
\left({M_x\over 1.4 \msun}\right)\left({T_{c}\over 10^{8} \K} \right)^{2.2}.
\end{equation} 
This luminosity will probably not quench the thermal instability of
the accretion disk as fully exposed matter at $R_r$ only has an
effective temperature of $T_{\rm eff}=1800 \K(T_c/10^{8}\K)^{0.55}$.
It is the unusual combination of large orbital separation, low
inferred mass transfer rates, and low neutron star luminosity that
allows for such instabilities in this neutron star binary.  For
time-averaged accretion rates $\approx 10^{-9} \msun \yr^{-1}$, the
outburst intervals are longer than the history of X-ray and
$\gamma$-ray monitoring. The system's extinction ($A_V>25$) also
makes optical identification of a prior outburst unlikely.

\section{
The Nuclear Burning of Accreted Matter
}

The thermal stability and appearance of nuclear burning on steadily
accreting neutron stars is well studied. For comparable metallicities
and magnetic fields weak enough so as not to affect the opacities
($<10^{13} \gauss$), the only residual difference between this neutron
star and others accreting at comparable local rates is a colder
core. This might allow for unstable hydrogen/helium ignition at
slightly higher instantaneous local accretion rates ($\dot m$) than on
a steadily accreting star.

Constant accretion at these $\dot M$'s conductively heats the neutron
star core, which attains an equilibrium temperature of
$T_c=(2\mbox{--}4)\times 10^8 \K$ in about $10^3\mbox{--}10^4 \yr$ by
balancing this heating with neutrino cooling (Ayasli \& Joss
1982). The time-averaged $\dot M$ in GRO~J1744-28 is comparable to
many X-ray binaries, whereas the instantaneous $\dot M$ can clearly be
much higher. The cycling of $\dot M$ leads to a colder equilibrium
core than that of a neutron star steadily accreting at the same
time-averaged $\dot M$. This is because the thermal timescale of the
deep ocean and crust (which is where substantial energy release occurs
as matter is forced through the electron capture boundaries [Haensel
\& Zdunik 1990]) is $\sim 10\mbox{--}100$ years, much longer than the
outburst duration. There is thus insufficient time to heat the crust
to a temperature profile favorable for sending a large luminosity
$L\approx (\dot{M}/m_p)(1\MeV)$ into the core. In addition, between
outbursts, the cold outer envelope conducts away the accumulated heat
in the deep ocean and crust. For temperatures below $T_c\sim
\ee{2}{8}\K$ the core cools radiatively between outbursts, much as a 
young neutron star still hot from birth.

\subsection{
Settling and Nuclear Burning of the Accreting Matter
}
 
As described in \S3, the neutron star is accreting the hydrogen-rich
envelope of an evolved giant.  Because the giant has already lost an
appreciable amount of mass, the matter presently being transferred was
most likely processed in the companion's interior during its main
sequence lifetime. This would increase the helium content. The
$\beta$-decay limitations and high temperatures on the neutron star
fix the hydrogen burning (via the CNO cycle) rate at the $\beta$-decay
limited value of the ``hot'' CNO cycle, $\epsilon_H=5.8\times 10^{15}
Z_{\rm CNO} \erg\gram^{-1}\secd^{-1}$ where $Z_{\rm CNO}$ is the mass
fraction of the CNO elements. For high accretion rates, this burning
never consumes the accreted hydrogen before helium ignition, so that
unstable helium burning occurs in a hydrogen-rich environment, 
which enhances the nuclear reaction chains and energy release (Lamb \& Lamb
1978; Taam \& Picklum 1979; Fujimoto et al.\ 1981; Taam 1982). Because
of the complicated rp process (Wallace
\& Woosley 1981), the exact composition of the ashes is unknown except
for a few limited cases (Van Wormer et al.\ 1994).

Prior to helium ignition, the accreted material is settling onto the
neutron star and is stably burning hydrogen. Since the thermal time is
always less than the time to accrete to a given depth, $t_{\rm
accr}=y/\dot m$, we find the temperature structure by solving the
time-independent entropy and flux equations, 
\begin{equation}\label{eq:atmos}
T\dot m{ds\over dy} ={dF \over dy }+ \epsilon_H,
\hskip 10 pt F={c\over 3 \kappa }{d a T^4\over dy}, 
\end{equation}
where the local accretion rate is written as $\dot m=\dot m_4 10^4
\gram\cm^{-2}\secd^{-1}$, $s$ is the specific entropy, $F$ is the
outward heat flux, and $y$ is the column depth.  The opacity in the
upper atmosphere is given by $1/\kappa=1/\kappa_{\rm
rad}+1/\kappa_{\rm cond}$, where the radiative opacity is the sum of
electron scattering (we use Paczy\'nski's [1983] fitting formulae
for the degeneracy and high temperature corrections) and free-free
absorption.  We use the conductivity from Yakovlev \& Urpin (1980).

These settling solutions are valid until the helium burns fast enough
to appreciably change either the temperature or the helium
abundance. Fushiki and Lamb (1987) defined the boundary of stable
helium burning in the $y$-$T$ plane (the ignition curve) by setting 
\begin{equation} 
{d\epsilon_{3\alpha}\over dT}={d\epsilon_{c}\over dT},
\end{equation}
where
$\epsilon_c=acT^4/3\kappa y^2$ is a local representation of the
conductive cooling and $\epsilon_{3\alpha}$ is the helium burning
rate. These derivatives are taken at constant pressure, as the
instability grows faster than the pressure changes. The heavy
dot-dashed line in Figure 2 denotes this ignition curve for
$Y=0.3$. Regions to the right of this curve are thermally unstable to
helium burning. We also define the depletion curve by equating the
helium lifetime to the $3\alpha$ reaction with $t_{\rm accr}$ (Fushiki
\& Lamb 1987); the heavy solid lines in Figure 2 denote this condition
for $\dot m_4=3,\ 7.5,\ {\rm and}\ 30$.

We derive the settling solutions by varying the flux exiting the
atmosphere until the flux is zero at the depth where the ignition
curve (or depletion curve, whichever is first) is met. We take zero
flux at the bottom to mimic the cold core. These settling solutions
all strike the ignition curve before the depletion curve, so that we
find the column density $y_{\rm ign}$ accumulated on the star prior to
unstable helium ignition.

For Population I type metallicities, the temperature of the settling
material is mostly set by the slight amount of hydrogen
burning. Indeed, as is evident in the three light solid lines that
display these settling solutions in Figure 2, $y_{\rm ign}$ depends
only mildly on $\dot m$ for the $Z_{\rm CNO}=0.01$ case. We obtain
$y_{\rm ign}/10^8 \gram\cm^{-2}=3.1,\ 2.95,\ {\rm and} \ 2.65$ for
$\dot m_4= 3, \ 7.5,\ {\rm and}\ 30$; these correspond to recurrence
times of 2.9 hours, 1.1 hours, and 15 minutes. The flux exiting the
atmosphere is $F/(10^{22}\erg\cm^{-2}\secd^{-1}) = 2.0, \ 2.42,\ {\rm
and}\ 4.93$ for these $\dot m$'s.

Lower CNO abundances are relevant if the companion is an older
Population II giant and/or if spallation of the incident nuclei occurs
in the accretion shock (Bildsten, Salpeter, \& Wasserman 1992).  At
lower $Z_{\rm CNO}$, the settling solutions are more sensitive to the
accretion rate via the gravitational compression terms.  The two
bottom thin solid lines in Figure 2 are settling solutions when
$Z_{\rm CNO}=10^{-4}$ for $\dot m_4=7.5\ {\rm and}\ 30$, giving
recurrence times of 6.6 hours and 0.5 hours.

\subsection{
When is the Burning Time Dependent?
}

The crucial question to answer is, ``At what accretion rate is the
burning unstable?''. The simplest way to address this is to construct
the steady-state solution (represented by the dotted lines in Figure 2
for the $\dot m$'s given above) that burns the material as fast as it
accretes. If this solution does not consume the fuel before reaching
the instability curve, then it is unstable.  For an accretion rate
below $\dot m_{\rm crit}\approx 3\times 10^4 \gram\cm^{-2}\secd^{-1}$,
the burning is unstable (for $Y=0.3$). This critical accretion rate
increases by a factor of two if the helium mass fraction is 0.5.
These estimates also indicate that the burning is stable when $\dot m
> \dot m_{\rm crit}$.\footnote{There have been indications in the past
that the $\dot m_{\rm crit}$ obtained from a time dependent simulation
might actually be higher than our estimate. In particular, Ayasli and
Joss (1982) argued that the steady state solution would not be reached
until the time to reach the ignition depth became shorter than the
local thermal time. This requires accretion rates 2--5 times higher
than our estimate. However, in the absence of a full time-dependent
calculation, we will stick to our present, and maybe overly conservative,
estimate.} {\em Since $\dot m_{\rm crit}$ is smaller than our minimum
estimated accretion rate at the outburst peak, it is unlikely that the
hourly bursts during the brightest parts of the outburst have a
thermonuclear origin.}

However, the burning becomes unstable when $\dot m< \dot m_{\rm crit}$
(either due to a reduction in the overall accretion rate or spreading
away from the polar cap; see \S5).  When the fuel spreads over the
whole star prior to ignition, the maximum luminosity (and hence flux)
at which an instability occurs is $L\approx 7\times 10^{37}
\erg\secd^{-1}$ (for $\dot m_{\rm crit}=3\times 10^4
\gram\cm^{-2}\secd^{-1}$ and $R=10 \km$).  This corresponds to 
a bolometric flux
\begin{equation}\label{eq:fmin}
F_{\rm unstable}<10^{-8} \erg\cm^{-2}\secd^{-1}\left(L\over 
7\times 10^{37} \erg \secd^{-1}\right)\left( 8 \,{\rm kpc} \over
d\right)^2. 
\end{equation} 
We convert this to $\cps$ in the PCA instrument aboard {\em RXTE\/} by
using the conversion factor (Giles et al.~1996) of $4\times 10^{-8}
\erg\cm^{-2}\secd^{-1}$ (2\mbox{--}60 keV) for $10^4\cps$. 
For $d=8 \,{\rm kpc}$, unstable burning signatures will most likely
not appear until the PCA count rate is $\lesssim 2500\cps$.

When the burning is unstable, the intersection of the settling
solutions with the helium ignition curve defines $y_{\rm ign}$.  For
typical metallicities, $y_{\rm ign}> \ee{1.5}{8}\gram\cm^{-2}$, which
at $\dot m_{\rm crit}$ gives a recurrence time of just over an
hour. Once the helium ignites, the temperature rises rapidly and
starts a propagating combustion front. If the whole star is ignited
(see next section) the maximum burst energy would be $4\pi R^2 y_{\rm
ign} (7 \MeV/m_p)$. This is the maximum, as time-dependent
calculations of the hydrogen/helium burning flash often found
incomplete burning (Taam et al.\ 1993) and the whole star need not
ignite. The resulting fluence would be
\begin{equation}
\mbox{Maximum Burst Fluence}=1.5 \times 10^{-6} \erg\cm^{-2} 
\left(y_{\rm ign}\over \ee{1.5}{8} \gram\cm^{-2}\right)\left(R\over
10\km\right)^2\left(8 \,{\rm kpc} \over d\right)^2.
\end{equation}
The peak luminosity depends on the combustion front's propagation
speed through the fuel-rich regions, as we now discuss.

\section{
The Role of the Magnetic Field
}

Conventional Type I X-ray bursts are not seen from highly magnetized
($B\gtrsim 10^{12} \gauss$) accreting X-ray pulsars. This was at
first surprising because they accrete at rates comparable to X-ray
burst sources that are not obviously magnetic. Joss and Li (1980) 
explained the lack of bursts by {\em stabilizing\/} the nuclear burning
with two mechanisms. For this pulsar, the relevant mechanism is the 
increased local accretion rate on the polar cap. The magnetic field
funnels the accretion onto the polar cap and confines the accretion
mound until the ignition pressure is reached.  These constraints are
typically satisfied for steadily accreting X-ray pulsars with $\dot{M}
\gtrsim 10^{-10} \msun \yr^{-1}$, as the fractional area of the polar
cap only needs to satisfy $A_{\rm cap}/4\pi R^2 \lesssim 0.01$. This
is well within the estimates obtained by either following the field
lines from the magnetospheric radius to the star (Lamb, Pethick, \&
Pines 1973),
\begin{equation}\label{cap}
A_{\rm cap} \approx 10^{11}\cm^2 \xi^{-1}
	\left(\frac{B}{10^{11}\gauss}\right)^{-4/7}
	\left(\frac{\dot{M}}{10^{-8} \ \msun \yr^{-1}}\right)^{2/7}
	\left(\frac{R}{10\km}\right)^{9/7}, 
\end{equation} 
or allowing the
matter to penetrate through the magnetopause via a Rayleigh-Taylor
instability and attach to field lines at smaller radii (Arons \& Lea
1976; Elsner \& Lamb 1977).  

The absence of Type I X-ray bursts
in {\em all\/} X-ray pulsars seems unlikely given the range in magnetic
field strengths and accretion rates. Bildsten (1995) suggested that,
even when the burning is thermally unstable, a strong magnetic field
will inhibit the rapid lateral convective motion needed for the
combustion front to ignite the whole star in a few seconds (Fryxell \&
Woosley 1982).  The field strength required to halt the convective
($\sim 10^6 \cm\secd^{-1}$) propagation of burning fronts is not
known.  Convection is potentially stabilized when $B^2 > 8\pi P$
(Gough \& Tayler 1966), which requires $B\gtrsim 7\times
10^{11}\gauss$ in the helium burning region.
\footnote{
Even lower fields might slow the burning fronts, as the sub-sonic
velocities ($v_c\sim 10^6 \cm\secd^{-1}$) implied by efficient
convection can only push around fields of strength $B^2<8\pi \rho
v_c^2$, or $B<10^9 \gauss$ at the helium ignition depth (Bildsten
1995). Observations will most likely tell us the outcome for magnetic
fields in the intriguing regime $\rho v_c^2 \ll B^2/8\pi \ll P\:
(10^9\gauss < B < 7\times 10^{11} \gauss )$.}
Most inferred dipolar field strengths
for accreting X-ray pulsars easily satisfy this constraint. However,
GRO~J1744-28 does not and hence is an especially intriguing candidate
for showing Type I X-ray bursts.  

If the lateral convective motion is
inhibited (or if there is too little fuel for convection to occur
[Bildsten 1993]) then the burning front propagates at the slower speed
set by heat transport (electron conduction and/or radiative transport,
depending on the depth) 
\begin{equation}
v_{\rm slow}\approx 80 + 200 
\left(\frac{y_{\rm He}-y_q}{\ee{4.5}{7} \gram\cm^{-2}}\right)
\cm\secd^{-1}, 
\end{equation}
where $y_{\rm He}$
is the local helium column density (in $\gram\cm^{-2}$) and
$y_q=\ee{1.35}{8} \gram\cm^{-2}$ is the minimum column density needed
for a pure helium burning front to propagate (Bildsten 1995). This
relation is for a pure helium atmosphere and is a reasonable lower
limit to the mixed hydrogen/helium burning case.  For the typical
$y_{\rm He}\approx \ee{2}{8} \gram\cm^{-2}$ where the instability sets
in, $v_{\rm slow}\approx 400 \cm\secd^{-1}$, so that the burning front
crosses a $10^5 \cm$ polar cap in $t_{\rm cross}\approx 4 \,{\rm
min}$.  In this case, the thermonuclear instability would appear as a
flare of a few minutes duration (potentially time symmetric) with a
luminosity set by the amount of accumulated fuel and the time to burn
all of it, $L_{\rm flare}\approx 0.05 L_{\rm accr}(t_{\rm accr}/t_{\rm
cross})$ (Bildsten 1995).  

\subsection{
The Spreading of Accreted Fuel
} 

The local accretion rate, $\dot m$, onto the polar cap would greatly
exceed $\dot m_{\rm sph}$ if accretion onto GRO~J1744-28 is via a
polar cap of the size given by equation (\ref{cap}) (this is not clear
given the small pulse fraction and nearly sinusoidal pulses seen by
Finger et al.\ [1996a]). Because the fate of the accreted material
depends on $\dot m$, knowing the depth at which accreted matter flows
laterally over the surface and reduces $\dot m$ to $\dot{m}_{\rm sph}$
is crucial. The magnetic Reynolds number of the flow is $4\pi \sigma
\ell v/c^2$, where $\ell$ is the length over which the magnetic field
varies, $\sigma$ is the electrical conductivity, and $v=\dot{m}/\rho$
is the downward flow velocity. Using the pressure scale height, $h =
P/\rho g$, as an estimate for $\ell$, we find that the magnetic
Reynolds number is $ \sim 100$ in the helium burning region ($P\approx
10^{22}\erg\cm^{-3}$, $T\approx\ee{5}{8}\K$). As a result, the
magnetic field is frozen into the matter at the ignition depth.

For a small polar cap, we find the pressure where the magnetic field
can no longer hold up the accretion mound by balancing the transverse
pressure gradient (given by the lateral extent of the polar cap,
$\ell_{\rm cap}\sim \sqrt{A_{\rm cap}} <R$) with a distorted field.
Following Hameury et al.\ (1983), we presume spreading occurs at the
depth where the field is distorted by a large angle.  To illustrate
this problem, consider an azimuthally symmetrical and poloidal
magnetic field $\vec{B}=(B_\cpi,0,B_z)$ in cylindrical coordinates
$(\cpi,\phi,z)$. The characteristic length in the $z$-direction is the
pressure scale height $h=P/\rho g$ and the characteristic length in
the $\cpi$-direction is $\ell_{\rm cap}$. The accretion flow distorts
the field from $\vec{B} = (0,0,B)$ to the perturbed configuration
$\vec{B} = (B_\cpi,0,B-\delta B_z)$. Equating estimates of $\vec{J}$
obtained from $\curl\vec{B} = 4\pi\vec{J}/c$ and the $\cpi$-component
of $-\grad P + \rho\vec{g} + \vec{J}\vcross\vec{B}/c = \vec{0}$, and
using $\divr\vec{B} = 0$ to obtain a relation between $\delta B_z$ and
$B_\cpi$, we have
\begin{equation} \label{eq:fd} 
\frac{B_\cpi}{B} \sim \frac{4\pi h P}{\ell_{\rm cap}B{}^2}.
\end{equation}
For a fully ionized H/He mixture with an ideal gas equation of state,
the pressure where $B_\cpi\sim B $ is
\begin{equation}\label{eq:spread} 
P_{\rm spread} \sim 10^{22} \erg\cm^{-3} 
	\left(\frac{A_{\rm cap}}{10^{11}\cm^2}\right)^{1/2}
	\left(\frac{B}{10^{10}\gauss}\right)^2
	\left(\frac{\ee{2}{8}\K}{T}\right).  
\end{equation}
This pressure defines a boundary ($y_s$ in column depth) where the
matter starts to spread laterally. We will presume that above $y_s$
the local accretion rate is $\dot{M}/A_{\rm cap}$ and that below
$y_s$, the local accretion rate is the spherical value $\dot{m}_{\rm
sph}$.

\subsection{ Global Behavior of the Nuclear Burning for GRO~J1744-28 } 

Having shown that the nuclear burning is unstable when $\dot m< \dot
m_{\rm crit}\approx 3\times 10^4 \gram \cm^{-2} \secd^{-1}$, we now
estimate $\dot m$ at the ignition depth in terms of the dipole
magnetic moment $\mu$ and global accretion rate $\dot M$ (we are
assuming for simplicity that no higher-order multipole moments are
present).  We show in Figure \ref{fig:parspace} the cases $\xi=1.0$
and $\xi=0.5$.  Within this parameter space, the first requirement is
that the magnetospheric radius be less than the co-rotation radius
(eq.~[\ref{eq:rm<rco}] with $R$ set to $10\km$) and is indicated by
the unhatched area.  The second relation, indicated by the dot-dashed
line in Figure \ref{fig:parspace}, is $y_s = y_{\rm ign}$.  Below this
line, the accretion flow spreads before ignition, thereby lowering the
local accretion rate to $\dot{m}_{\rm sph}$. Depending on the dipolar
field strength there are two cases.
\begin{itemize} 

\item For $B\lesssim \ee{(2\mbox{--}4)}{10}\gauss$, the accreted matter
spreads before igniting regardless of the polar cap area.  Thus, the
relevant local accretion rate (indicated by the vertical dashed line)
is $\dot{m}_{\rm sph}$ and the burning is unstable when the flux
constraint (eq.~[\ref{eq:fmin}]) is reached. This case is denoted by
region I in Figure 3.

\item For $B\gtrsim \ee{(2\mbox{--}4)}{10}\gauss$, the matter ignites prior
to spreading, so that the nature of the thermonuclear burning depends
on the polar cap size. This case is denoted by region II in Figure
3. The most constraining polar cap area is given by equation \ref{cap}
and is indicated by the slanted dotted line in the right-hand plot.
For GRO~J1744-28, this scenario is only relevant if $\xi\sim 0.5$, as
otherwise the propeller effect will halt accretion before
$\dot{m}<\dot{m}_{\rm crit}$ on the polar cap.  For $\xi=1.0$, nuclear
burning in region II is unstable only if the polar cap area is larger
than the estimate of equation (\ref{cap}).

\end{itemize}

Our ignorance of the polar cap size prohibits us from saying at what
accretion rate the $B\gtrsim\ee{(2\mbox{--}4)}{10} \gauss$ case
becomes unstable (the triangular regions denoted II). If we choose a
fixed polar cap area of 10\% of the stellar area, then the $\dot M$
needed for an instability decreases by a factor of ten, as does the
maximum flux required to see an instability. The polar cap of Arons and
Lea (1980) is always larger than 10\% of the stellar area and implies
that the burning becomes unstable when $L< \ee{3}{37} \erg\secd^{-1}$.
The appearance of unstable burning will constrain the polar cap size,
as a necessary condition for stable burning is $A_{\rm
cap}<\dot{M}/\dot{m}_{\rm crit}$.

\section{
Summary and Observational Outlook
}

The continuous monitoring of GRO~J1744-28 by the {\em RXTE\/} provides
an important opportunity to learn about both accretion and
thermonuclear instabilities on a weakly magnetized neutron star. We
have shown that the bursts observed during the peak of the outburst
are most likely not of thermonuclear origin, as even the minimum local
accretion rate on the neutron star is too high for unstable burning of
hydrogen-rich material. This statement is no longer true as the
accretion rate, $\dot M$, decreases.

The full understanding of the thermonuclear burning depends on many
properties of both the neutron star and the binary.  In \S2, we used
the observed torque to estimate the minimum accretion rate and dipole
field, which led to the comparison with SMC X-1, another bursting
X-ray pulsar. We speculated in \S3.2 that the present outburst might
be the result of a thermal instability in the disk, comparable to what
occurs in dwarf novae. We also discussed the optical companion and
pointed out that a careful velocity measurement of the companion will
constrain the neutron star mass.

We have shown that, if $B\lesssim\ee{(2\mbox{--}4)}{10} \gauss$, then the
nuclear burning becomes unstable when the intrinsic source luminosity
is $L\lesssim \ee{7}{37} \erg\secd^{-1}$. For higher magnetic fields,
the matter stays confined at the polar cap before ignition, in which
case the stability of the nuclear burning depends strongly on the
polar cap size. As the outburst fades and $\dot M$ decreases, the
burning might become unstable, especially if the polar cap is larger
than the conventional estimate (equation [\ref{cap}]), for which the
thermonuclear burning is always stable when $\xi=1$. Hence, for larger
fields, the global accretion rate when the first signatures of
thermonuclear instability appear will constrain the polar cap size.

As discussed in \S5, the character of the instability strongly depends
on the burning front's propagation speed. If the instability is too
weak to convect (or if the field is strong enough, $B\gtrsim 7\times
10^{11} \gauss$, to inhibit the lateral convective heat transport) a
possible burning signature is flares with durations of a few minutes
to an hour. The duration of the flare is set by the slow burning speed
(\S5) and the size of the fuel-rich region. These flares rise on a
timescale comparable to their duration and are not necessarily
asymmetrical in time, as Type I X-ray bursts are.  For a polar cap
radius of $10^5\cm$ (10\% of the stellar radius) and a typical
ignition depth of $\ee{2}{8}\gram\cm^{-2}$, the burning front crosses
the polar cap in four minutes. If convection occurs, then we expect
to see Type I X-ray bursts as in other Low Mass X-Ray Binaries.

When bursts or flares first appear, they will have recurrence times of
order an hour if our $\dot m_{\rm crit}$ determination is accurate. It
is possible that the recurrence times will be longer than this
(especially for low metallicity), in which case detection will be more
difficult. The ability of the PCA to position bursts to within $0.2$
degrees has already eliminated GRO~J1744-28 as the source of two Type
I X-ray bursts seen in the field (Corbet \& Jahoda 1996; Jahoda et
al.\ 1996) and should eventually provide an unambiguous localization
of a Type I burst from GRO~J1744-28.

We thank Jon Arons, Mark Finger, Keith Jahoda, Chris McKee, Ed
Morgan, Tom Prince and Bob Rutledge for many discussions about the
nature of this source. We especially thank Bob Rutledge (MIT) for
creating and maintaining a Web site about this X-ray binary. Our work
was supported by NASA via grants NAG 5-2819 and NAGW-4517 and by the
California Space Institute (CS-24-95).  L. B. was also supported by
the Alfred P. Sloan Foundation.

\begin{figure}[ht]
\centering{\epsfig{file=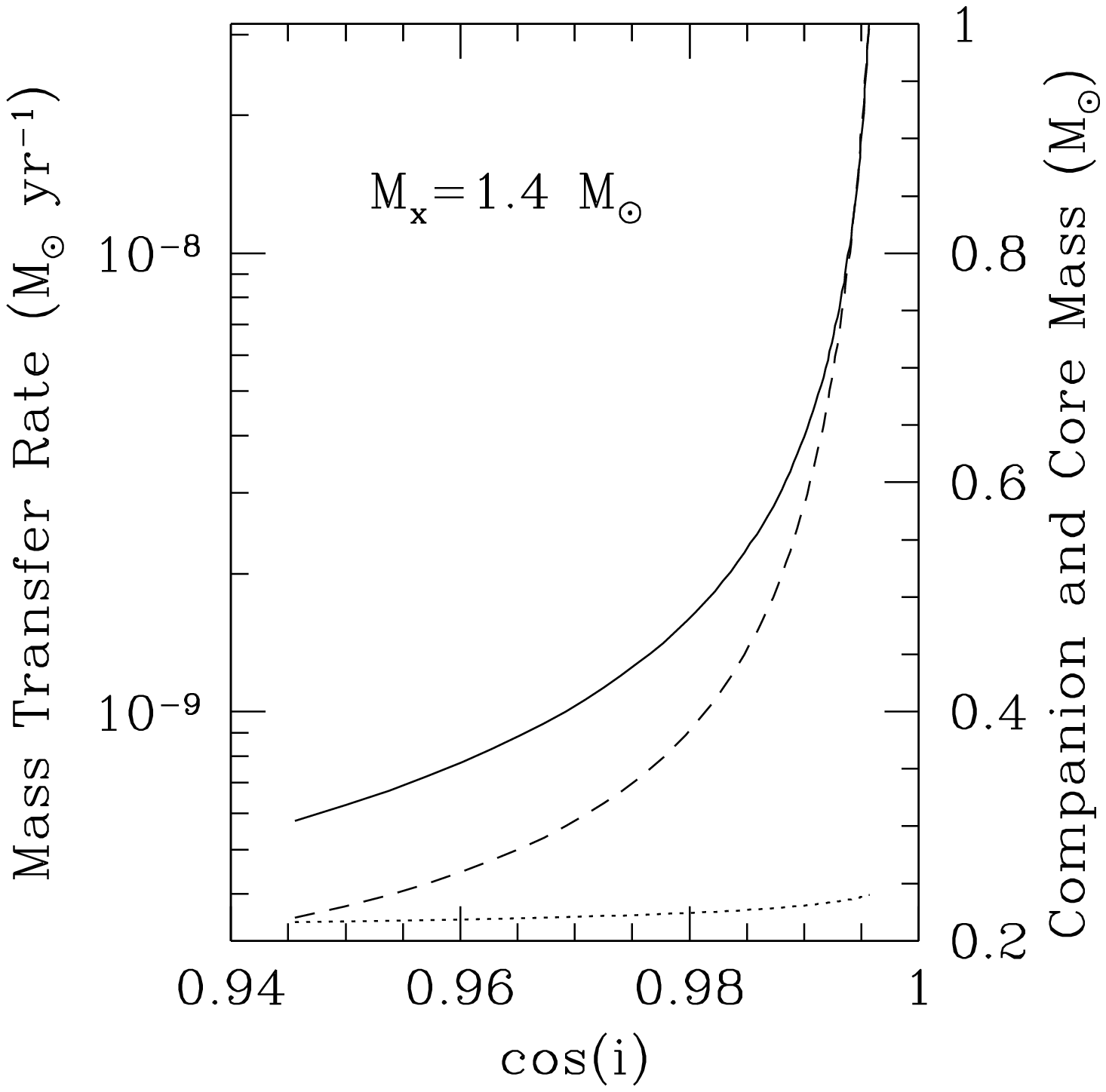,angle=0,height=6in}}

\caption{
The companion mass and mass transfer rates as a function of
inclination angle for $M_x=1.4 \msun$.  For the orbital parameters of
Finger et al.~(1996), we show the red giant companion mass (dashed
line) and inferred mass transfer rate (solid line) as a function of
$\cos i$. There are no Roche-lobe filling solutions for $i>18$ degrees
if $M_x=1.4 \msun$.
\label{fig:compan}}
\end{figure}

\begin{figure}[ht]
\centering{\epsfig{file=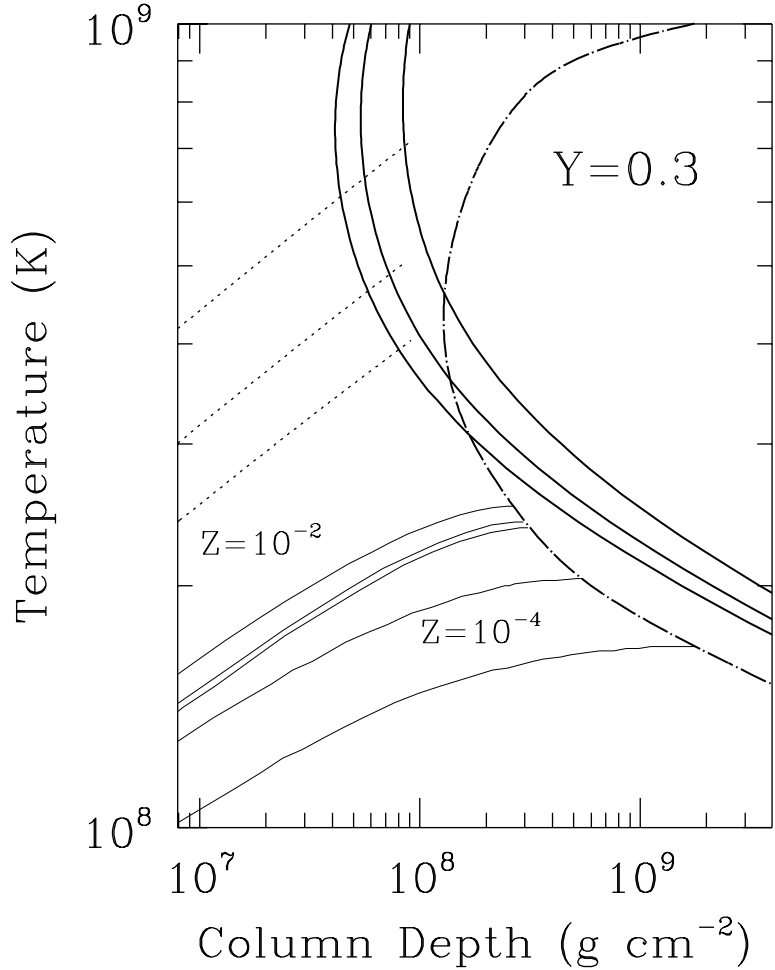,angle=0,height=6in}}

\caption{
Helium ignition for hydrogen-rich accretion ($Y=0.3$) at accretion
rates appropriate for this source for a $1.4 \msun$, $R=10
\km$ neutron star. The heavy solid lines show the helium depletion
curves for $\dot m_4=3,\ 7.5,\ {\rm and}\ 30$. The heavy dot-dashed
curve is the helium ignition curve.  The top three thin solid curves
are settling solutions for $Z_{\rm CNO}=0.01$ and $\dot m_4= \ 3, \
7.5,\ {\rm and} \ 30$, while the bottom two are for $Z_{\rm
CNO}=10^{-4}$ and $\dot m_4=7.5\ {\rm and}\ 30$.  The three dashed
lines are the envelope structure if the burning were in steady-state
for these three $\dot m$'s.}
\end{figure}

\begin{figure}[ht]
\centering{\epsfig{file=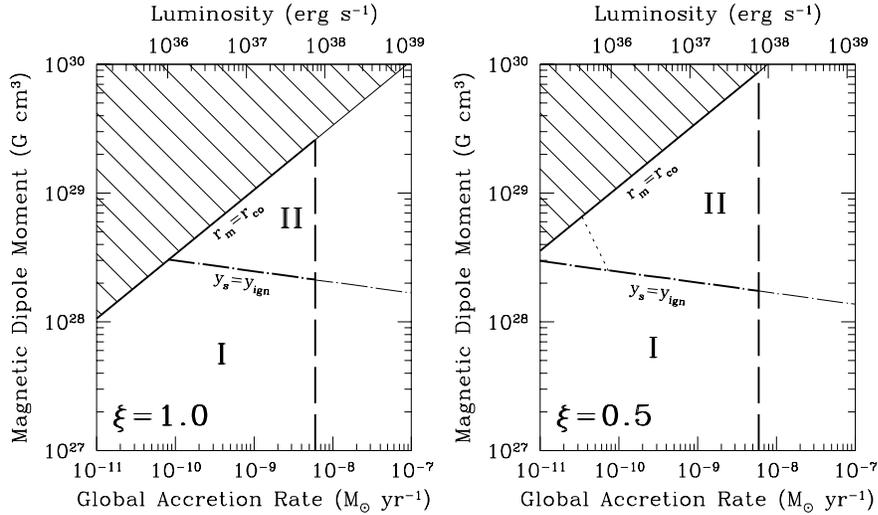,angle=0,height=5.7in}}

\caption{
A graphical summary of the burning behavior parameterized by the
dipole magnetic moment $\mu$ and the global accretion rate $\dot{M}$.
The left plot is for $\xi=1.0$; the right, for $\xi=0.5$
(cf.~equations [\ref{eq:rm<rco}] and [\ref{cap}]).  We assume for
simplicity that no higher-order multipole moments are present.  In the
shaded region the propeller effect stops accretion (i.e., $r_m>r_{\rm
co}$).  For the region beneath the dot-dashed line labeled $y_s =
y_{\rm ign}$, spreading occurs above the ignition column depth $y_{\rm
ign}\approx \ee{2}{8}\gram\cm^{-2}$.  The vertical dashed line
indicates the critical accretion rate, $m_{\rm crit} =
\ee{3}{4}\gram\cm^{-2}\secd^{-1}$, for spherically symmetric
accretion.  In region I, the burning is unstable regardless of polar
cap size.  For region II, the accreted material is magnetically
confined at the ignition column density; our ignorance of the polar
cap size prohibits us from indicating where the burning is unstable.
In the right-hand plot we show with a dotted line where unstable
burning begins if the accreted material is confined to a polar cap of
area given by equation (\ref{cap}). Nuclear burning is unstable in the
region to the left of this line.
\label{fig:parspace}}
\end{figure}

\end{document}